\documentstyle[aasms4]{article}

\begin{document}

\title{Small Scale Fluctuations of the Microwave Background in the Quasi-Steady State Cosmology}

\author{J.V. Narlikar, R.G. Vishwakarma, \affil{Inter-University Centre for Astronomy and
Astrophysics, Post Bag 4, Ganeshkhind, Pune 411 007, INDIA} G. Burbidge \affil{Center
for Astrophysics and Space Sciences, University of California, San Diego, CA 92093-0424, USA} and F. Hoyle,
\affil{102 Admirals Walk, West Cliff, Bournemouth, England, UK}}

\begin{abstract}

In this paper we show that, within the framework of the QSSC, the small scale 
deviations on angular scales $\lesssim 1^{0}$ expected in the MBR are due to 
inhomogeneities in the distribution of galaxies and clusters. It is shown how these can be estimated
on the galaxy-cluster-supercluster scale at the epoch of redshift $\sim 5$ when the universe was last passing through the minimum phase of the scale factor.  Rich clusters on the scale of 5-10 Mpc generate the kind of peak in the fluctuation power spectru
m observed by the Boomrang and Maxima projects.  Weaker inhomogeneities on smaller scales with $l~\sim 10^3$ are expected to arise from individual galaxies and small groups.

\end{abstract}


\section{INTRODUCTION}

The quasi-steady state cosmology (QSSC) has been proposed by Hoyle, Burbidge and 
Narlikar (1993, 1994$a$, $b$, 1995) as an alternative to the standard big bang model.  This cosmology does away with the initial singularity, and does not have any cosmic epochs when the universe was hot. The synthesis of light nuclei and the origin of th
e microwave background radiation have therefore to be  shown to be explained by different physical processes than those invoked in the 
big bang. (cf. Hoyle, Burbidge and Narlikar, 2000; Burbidge and Hoyle, 1998].  

The microwave background radiation (MBR) is explained as follows.  A typical QSSC scale factor for a flat Robertson-Walker universe is given by

\begin{equation}
S(t) = {\rm exp} (t/P) \times [{1~+~\eta {\rm ~cos~} 2\pi \tau/Q}],
\end{equation}

\noindent where the time scales $P~\sim 10^{12}$ yr $\gg$ $Q~\sim 40-50 $ Gyr are considerably greater than the Hubble time scale of 10-15 Gyr of standard cosmology. The function $\tau(t)$ is very nearly like $t$, with significantly different behaviour ne
ar the minima of the function $S(t)$. The parameter $\eta$ has modulus less than unity thus preventing the scale factor from reaching the value zero.  Hence there is no spacetime singularity nor a violation of the law of conservation of matter and energy,
 as happens at the big bang epoch in the standard model. For, matter in the universe is created through minibangs, through explosive processes that produce matter and an equal quantity of a negative energy scalar field $C$.  Such processes take place when
ever the energy density of the $C$-field rises above the threshold energy $m_P c^2$, the restmass energy of the Planck particle which is typically created.

While the overall level of the $C$-field is below this threshold, it can rise in the neighbourhood of highly collapsed massive objects (objects close to becoming black holes), and these then become the sites of creation.  Also since the overall energy den
sity of the $C-$field goes as $S^{-4}$, the creation is facilitated near the epochs of smallest $S$.  It is close to these relatively denser epochs that the creation activity is at its peak, switching itself off as the universe continues to expand in the 
cycle.  Hence from one cycle to another the density of matter would have fallen by the factor exp$(-3Q/P)$, but for the compensating creation of matter at the beginning of a new cycle at the minimum of $S$.

Thus, although the universe is in a long term steady state, it oscillates over shorter time scales, with each cycle physically the same as the preceding one.  Each cycle has new matter forming into stars and galaxies which evolve as in the previous cycle.
  What happens to the radiation of stars in this process?  Assuming that the radiation density falls like $S^{-4}$ in each cycle, it would be depleted by the factor exp$(-4Q/P)$ from one oscillatory minimum to the next.  It is this gap that is made up by t
he starlight generated during the cycle.  Given  $P$ and $Q$, and the starlight energy generated per cycle, we can estimate the radiation background at any stage of a cycle.   
We have shown elsewhere (Burbidge and Hoyle 1998; Hoyle, Burbidge and Narlikar 2000) that
it is this starlight thermalized by dust grains which is responsible for the microwave background
(MBR) in the QSSC.  We have also shown why the temperature is very close to $2.7^0$ K.
In the following sections we show how the processes of thermalization take place. 

In the following sections we show how the smoothness, the spectrum and homogeneity of the MBR are explained in the QSSC and compare the results with recent observations in which the 
fluctuations of the MBR at different angular scales have been measured.

\section{SPECTRUM AND HOMOGENEITY OF THE MBR}

In our model it is hydrogen burning in stars in galaxies that generates the energy of the MBR,
and the energy distribution of stellar radiation over different wavelengths from a typical galaxy varies through the cycle in the following way.  At the minimum $S$ epoch, new gaseous material is acquired and new stars are formed, with the energy coming ou
t mostly in the blue to ultraviolet, from stars of masses greater than $M_{\odot}$.  However, as the cycle proceeds towards the maximum phase and the times of the order $\sim 25$ Gyr elapse, the massive stars burn out, leaving only the low mass stars stil
l shining.  The typical stars left are the dwarfs of types $K$ to $M$, with the consequence that the radiation is then mainly in the red and the infrared.  However, as the cycle proceeds towards the next minimum, these wavelengths are shortened  as the sc
ale factor drops by a considerable factor.  

To fix ideas, we will consider the typical oscillatory cycle from one minimum of scale factor to next. Let 

\begin{equation}
x = S(t)/S(t_{min})
\end{equation}

\noindent where, in going from one minimum to the next we have ignored the exponential term. It is clear from this time dependence, that in going from the maximum to the minimum the scale factor drops by a factor $(1~+~\eta)/(1~-~\eta) \sim 12$ for $\eta 
\sim 0.85$.  We shall assume as an additional parameter that the present epoch is characterized by $x~=~6$.

Thus, infrared wavelengths of 1000 nm at $x=6$ are shortened to ultraviolet wavelengths of only $\sim 160$ nm at $x=1$.  However, by this stage, the intergalactic density is increased by a factor $x^{-3}$ as the minimum is approached and the issue of abso
rption becomes important.

We have shown that metallic particles in the form of graphite whiskers as well as iron whiskers play a crucial role in the absorption process. As Narlikar et al (1997) have discussed, a plausible case can be made for the condensation of metallic whiskers 
from the hot ejecta of supernovae, which are blown out into the intergalactic space by the shock waves generated at the time of explosion. 
The extinction properties of such whiskers, typically of length $\sim 1$mm and diameter $0.01\mu$m differ considerably from those of normal spherical dust.  In particular the iron whiskers at cryogenic temperatures are a dominant source of opacity in the 
microwave region, while carbon whiskers are more effective at the shorter UV wavelengths.  

The value of $Q_{{\rm abs}}$ for graphite whiskers is essentially constant for all wavelengths longer than $\sim 1 \mu$m, extending even to the long radio wavelengths, being equivalent to an absorption coefficient of $10^5$cm$^2$g$^{-1}$.  It is, however,
 three times this value for the UV radiation.  Thus an intergalactic density of $\sim 10^{-34}$ g cm$^{-3}$ at $x=6$ would rise to $\sim 2\times 10^{-32}$g cm$^{-3}$, at the minimum ($x=1$) epoch.  Over a cosmological distance of $10^{27}$cm at this stage
, the optical depth would be $\sim 6$ for UV, and $\sim 1$ for wavelengths longer than 1 $\mu$m.

The great bulk of the optical radiation that becomes subject to thermalization in the contracting phase of the oscillation will have travelled a distance of the order of 10 Gpc, or even more in the case of microwaves. The radiation incident on a carbon wh
isker will have been in transit since the maximum $S$ epoch of the previous cycle, and it includes all the microwave radiation existing {\it before} the present cycle, as well as the starlight generated by the galaxies in the current cycle.  The result is
 that all this radiation is well thermalized and uniform in energy density. The carbon whiskers themselves can, however, be lumpily distributed, on the scale of clusters of galaxies.  So as the minimum of $S$ is approached at the end of the last cycle, th
e conversion of starlight to microwaves would be lumpy. Let us consider a typical carbon whisker reacting to the ambient radiation.

To begin with, the whisker will find itself in a uniform radiation bath, which includes microwaves from previous cycles as well as starlight still to be converted to them.  As the starlight is progressively absorbed, the grain temperature first rises abov
e the MBR temperature $T_{{\rm MBR}}$ to a value

\begin{equation}
T_g\sim~[(1-f)~+~3^{1/4}\times f]\times~T_{{\rm MBR}}
\end{equation}

\medskip

\noindent where we have assumed that a fraction $f~\sim 10$ percent of the radiation is in the ultraviolet having been blueshifted during the contracting phase, so that the stellar component  has a $Q_{abs}$ three times higher.  With optical depth $\tau$ 
for the progressive absorption of the starlight, the grain temperature drops according to the relation

\begin{equation}
T_g\sim~[ 1~+~f\times {\rm e}^{-\tau}\times (3^{1/4}~-~1)]\times~T_{{\rm MBR}}.
\end{equation}

\noindent Thus the grain temperature lapses back to the temperature of the microwave 
background. 

The effect of grains being lumpily distributed on the scale of clusters of galaxies on the MBR is to cause the slight rise followed by the fall back to the MBR temperature to be lumpy too. The radiation background itself, however, does not derive this lum
piness: for the total assembly of grains has a negligible heat content and in thermal equilibrium, each one emits as much heat as it receives.  Since the emissivity of the particles does not depend on wavelength, each emits a Planckian spectrum correspond
ing to its temperature $T_g$.  Eventually when all particles come down to the temperature $T_{{\rm MBR}}$, however, further absorption and reemission bring about a strict Planckian distribution at this temperature.  Although as suggested by Narlikar, Wick
ramasinghe and Edmunds (1975) the thermalization can be achieved by the carbon whiskers, the presence of a small quantity of iron whiskers helps the process further.  The absorptions and reradiations by the whiskers at the oscillatory minimum, thus genera
te a mixing of radiation from distances as far as $\sim 10^{29}$ cm, at the relatively low intergalactic particle densities prevalent there, thus permitting radiation to travel freely.

\section{THE ORIGIN OF MBR INHOMOGENEITIES}

Although at the present epoch, the last minimum was some 10 Gpc away, there will be local sources, such as a galaxy at  a distance of $d \sim$ 3 Mpc, which are capable of making a small contribution to the MBR.  Galaxies in which the main contribution to 
stellar radiation comes essentially from the main sequence dwarfs, would have a luminosity of $L \sim 10^{44}$ erg s$^{-1}$.  At a distance of 3 Mpc, the energy density of its radiation would be around  

\begin{equation}
\epsilon = \frac{L}{4\pi d^2}\times c \sim 3\times10^{-18} {\rm erg cm}^{-3},
\end{equation}

\noindent which may be compared to the energy density of starlight of the order of $10^{-14}$ erg cm$^{-3}$, i.e., a local variation of energy density of one part in 3000. This leads to a variation in temperature of the thermalizing particles by one part 
in $\sim 12,000$.  This effect becomes negligible when we take into account all thermalizing particles out to the cosmological distances.
However, when we take into account a particle lying within $\sim 3$ Mpc of a rich cluster containing, say, 1000 galaxies, the local modification of the starlight energy density will be of the order $1000\times 3\times 10^{-18}$ erg cm$^{-3}$ = $3\times 10
^{-15}$ erg cm$^{-3}$.  This is around 30 percent of the average starlight energy density, and is thus capable of changing the temperature of the typical whisker from the ambient MBR temperature by an amount that is 30 percent of the value calculated in e
quation (3).  For the present temperature of 2.73 K, this is $\sim 3\times 10^{-2}$ K.

Consider now the sky examined on an angular scale such that a beam width containing a rich cluster is compared with one that does not contain such a cluster. Based on the above calculation, shall we expect to find a detectable variation of background temp
erature between the two observations?  The answer is, `yes', but we have to take into account the local fluctuation along with all the thermalizing particles out to the Olbers limit $\sim c/H_0, ~H_0$ being the present value of the Hubble constant. This d
istance being of the order of 5000 Mpc, when compared with the the local distance of 3 Mpc, the overall effect will be down to a temperature variation of the order of 3/5000 of the above estimate of  $3\times 10^{-2}$, K, i.e., about 20$\mu$K.

We can look at the calculation another way.  Imagine a 10 Mpc size region, typically containing $\sim 10^4$ galaxies, each emitting starlight at the rate $\sim 10^{44}$ erg /s ( corresponding to magnitude -21.4). The flux of radiation in the form of degra
ded starlight across the surface of this region (assumed to be a sphere of radius 5 Mpc), would be $\sim 3.35 \times 10^{-4}$ erg/s/cm$^2$. Now, the present energy density of MBR $\sim 4.2 \times 10^{-13}$ erg cm$^{-3}$, implies that by the $(1+z)^{-4}$ r
ule, it was $6^4$ times this value at the epoch of the last minimum of $S$.  Using the Stefan's law the flux across a sphere of radius 5 Mpc for this radiation would be $c/4$ times the energy density, i.e., about 4.08 erg/s/ cm$^{2}$.  Comparing the exces
s flux due to the galaxies in this region computed above with this value we find that the excess flux is by a factor $8.21\times 10^{-5}$ of the average background.  Equating this to $4\Delta T/T$, we get the fluctuation of temperature as $\Delta T~=~56 \
mu K$.  Given the observational uncertainties
this compares well with the peak recently given by the Boomerang results (de Bernardis et al 2000; Hanany et al 2000).

Because of the inhomogeneity of the universe on the scale of clusters, the calculation cannot be made more precise, and at this stage one should only look at its order of magnitude.  The fact that this calculation yields a temperature fluctuation of the o
rder reported by the various observations is, however, encouraging from the point of view of the QSSC.  We now estimate the angular scale of the inhomogeneities.

\section{THE ANGULAR SCALE OF INHOMOGENEITIES}

Typically we will look at the angle subtended by a linear scale $a$ of an object located at redshift $z$ in the quasi-steady state model whose scale factor is defined in equation (1).  To fix ideas we will choose the parameters which we have also used for
 explaining the redshift magnitude relation based on Type Ia supernovae ({\it see} Banerjee, et al, 2000).  They are :

\begin{equation}
H_0 =65 {\rm ~km s}^{-1}{\rm Mpc}^{-1}, ~~~, z_{{\rm max}} = 5, ~~~, \eta~=~0.811, ~~~\Lambda_0~=~-0.358.
\end{equation}

The parameters $P$ and $Q$ do not enter explicitly in the calculation but we may typically take them to be 1000 Gyr and 50 Gyr, respectively.  
We further define the dimensionless parameters by the following formulae:

\begin{eqnarray} 
\Omega_0&=&\frac{8\pi G\rho_0}{3H_0^2}~~~{\rm density~  parameter},\nonumber \\
\Lambda_0&=&\frac{\lambda}{3H_0^2}~~~
{\rm cosmological~ constant~ parameter}, \nonumber
\\ \Omega_{c0}&=&\frac{8\pi G\rho_{c0}}{3H_0^2}~~~{\rm creation~ density~ 
parameter}, \nonumber \\ 
K_0&=&\frac{k}{H_0^2S_0^2}~~~{\rm curvature~ parameter}.
\end{eqnarray}

\noindent Here $\lambda$ is the cosmological constant, which is negative in the QSSC. The angle subtended at the observer by the above scale at redshift $z$ ($see$ Banerjee and Narlikar, 1999) is then given by

\begin{equation}
\alpha = \frac{H_0a(1~+~z)}{c}\int^{1+z}_{1}\frac{cdy}{\sqrt{\Lambda_0~+~\Omega_0y^3~+~\Omega_{C0}y^4}}.
\end{equation}

\noindent We will take $z~=~z_{{\rm max}}~=~5$.  Then the above calculation gives

\begin{equation}
\alpha~=~3a.
\end{equation}

\noindent where $a$ is measured in units of megaparsecs.

\noindent This then is the characteristic angle subtended by an object of linear size $a$ viewed when close to the oscillatory minimum.  Assuming that this is the size of a rich cluster of galaxies, we expect the characteristic angular size of the MBR inh
omogeneity it generates will be of this order.  This would
translate to a peak in the angular power spectrum at a harmonic of order
$l ~\sim 180^{0} / \pi \alpha$. So we arrive at the following relationship between cluster size and the harmonic it generates:

\begin{equation}
l~\cong \frac{1100}{a}.
\end{equation}

\noindent Thus a peak in the power spectrum of MBR anisotropy at harmonic $l~\sim 200$ as observed by the Boomerang and the Maxima groups 
(de Bernardis et al 2000, Hanany et al, 2000)  is generated in this theory by rich clusters of size $\sim 5.5$ Mpc, located at cosmological distances.  Smaller peaks are expected on the scale of groups of galaxies, at $l~\sim 500-1000$, and on the scale o
f superclusters at lower values of $l~\sim 10-20$.  These latter peaks will be smaller than the cluster peak at $l ~\sim 200$ and
thus they will be much harder to detect.

\section{CONCLUDING REMARKS}

In the QSSC the MBR arises from the matter in galaxies and this is quite different from its origin
in big bang cosmology.  In a similar way the major peak in the MBR in the framework of the
QSSC is explained in terms of rich clusters of galaxies, whereas in the big bang cosmology
it is the Doppler peak. The QSSC interpretation links the inhomogeities of radiation field to those of the matter field, of visible matter in the form of clusters of galaxies.  Indeed, one may argue that in the QSSC interpretation, the observations of  MB
R-inhomogeneities provide us with a direct diagnostic of the structural hierarchy in the universe.  For example, the pattern of patchiness could be related to the cluster - cluster correlation function.

It should be also mentioned that the interpretation of the $m-z$ relation for Type Ia supernovae
within the framework of the QSSC by Banerjee et al (2000) invokes the same whisker - dust as is used here for
generating a redshift - dependent contribution to $m$.  It is significant that the dust density
required to get  the best (and highly satisfactory) fit to the observed $m-z$ relation agrees
very well with that needed for generating the MBR.

\acknowledgements
We thank Tarun Souradeep for comments. J.V. Narlikar and R.G. Vishwakarma acknowledge grants from the Department of Atomic Energy for support of the Homi Bhabha Professorship for the former and the associated postdoctoral fellowship for the latter.

\newpage

\centerline {{\bf References}}
\medskip

\noindent Banerjee, S.K. and Narlikar, J.V. 1999, {\it MNRAS}, {\bf 307}, 73
\medskip

\noindent Banerjee, S.K., Narlikar, J.V., Wickramasinghe, N.C., Hoyle, F. and Burbidge, G. 2000, {\it A.J.}, {\bf 119}, 2583
\medskip

\noindent Burbidge, G. and Hoyle, F. 1998, {\it Ap. J.}, {\bf 509}, L1
\medskip

\noindent de Bernardis, et al, P. 2000, {\it Nature}, {\bf 404}, 955
\medskip
 
\noindent Dittmar,W. and  and Neumann, K. 1958, in {\it Growth and Perfection in Crystals}, Eds R.H. Doremus, P.W. Roberts and D. Turnbull, Wiley,388
\medskip

\noindent Gomez,R.J. 1957, {\it Chem. Phys.}, {\bf 26}, 1333
\medskip

\noindent Hanany, S. et al 2000, {\it Ap. J.}, {\bf 545}, L5
\medskip

\noindent Hoyle,F., Burbidge, G. and Narlikar, J.V. 1993, {\it Ap. J}, {\bf 410}, 437
\medskip

\noindent Hoyle,F., Burbidge, G. and Narlikar, J.V. 1994$a$, {\it MNRAS}, {\bf 267}, 1007
\medskip

\noindent Hoyle, F., Burbidge, G. and Narlikar, J.V. 1994$b$, {\it Astron. \& Ap.}, {\bf 289}, 729
\medskip

\noindent Hoyle,F., Burbidge, G. and Narlikar, J.V.  1995, in {\it The Light
Nuclear Abundances}, Ed P. Crane, Springer Verlag, p.21
\medskip

\noindent Hoyle,F., Burbidge, G. and Narlikar, J.V. 2000, {\it A Different Approach to Cosmology}, Cambridge : Cambridge University Press
\medskip

\noindent Nabarrow, F.R.N. and Jackson, P.J. 1958, in {\it Growth and Perfection in Crystals}, Eds R.H. Doremus, P.W. Roberts and D. Turnbull, Wiley, 65 
\medskip

\noindent Narlikar,J.V., Wickramasinghe, N.C., Sachs, R. and Hoyle, F. 1997, {\it I.J. Mod. Phys}. {\bf D6}, 125
\medskip

\noindent Narlikar, J.V., Wickramasinghe, N.C. and Edmunds, M.G. 1975, in {\it Far Infrared Astronomy}, Ed. M.Rowan-Robinson, Pergeman, 131

\end{document}